\documentstyle[12pt,epsf]{article}

\makeatletter
\@addtoreset{equation}{section}

\newcommand{\what}{\widehat}

\newcommand{\nn}{\nonumber}
\newcommand{\vs}[1]{\vspace*{#1}}
\newcommand{\hs}[1]{\hspace*{#1}}
\newcommand{\tr}{\mathop{\rm Tr}}
\newcommand{\p}{\partial}
\newcommand{\Half}{\frac12}
\newcommand{\unit}{\hbox to 3.8pt{\hskip1.3pt \vrule height 7.4pt
    width .4pt \hskip.7pt \vrule height 7.85pt width .4pt \kern-2.4pt 
    \hrulefill \kern-3pt \raise 3.7pt\hbox{\char'40}}}

\def\href#1#2{#2}

\begin{document}

\vspace{10mm}
\begin{titlepage}
\title{
\hfill\parbox{4cm}
{\normalsize KUNS-1642\\{\tt hep-th/0002090}}\\
\vspace{1cm}
Branes and BPS Configurations of\\
Non-Commutative/Commutative Gauge Theories
}
\author{
Koji {\sc Hashimoto}\thanks{{\tt hasshan@gauge.scphys.kyoto-u.ac.jp}}
{} and
Takayuki {\sc Hirayama}\thanks{{\tt hirayama@gauge.scphys.kyoto-u.ac.jp}}
\\[7pt]
{\it Department of Physics, Kyoto University, Kyoto 606-8502, Japan}
}
\date{\normalsize February, 2000}
\maketitle
\thispagestyle{empty}

\vs{10mm}

\begin{abstract}
\normalsize\noindent 
We study BPS Dirac monopole in $U(1)$ gauge theory on non-commutative
spacetime. The corresponding brane configuration is obtained in the
equivalent ordinary gauge theory through the map proposed by Seiberg
and Witten. This configuration coincides exactly with a tilted
D-string as predicted. This study provides an interesting check of the 
equivalence of the non-commutative and ordinary gauge theories.
\end{abstract}

\end{titlepage}

\section{Introduction}

In these five years, string theory has provided various interesting
tools for analyzing field theories \cite{Hanany:1996ti, Witten:1997so, 
Maldacena:1997tl}. This owes to the fact that 
effective theories on D-branes, non-perturbative solitons in
string theory, are supersymmetric gauge theories in various
dimensions \cite{Leigh:1989}. Since string theory contains lots of
perturbative and
non-perturbative dualities \cite{Hull:1994}, consequently various
field theories are
related by the string dualities. Through string theory, one obtains
non-trivial equivalence between different field theories.

One of the most intriguing examples is the equivalence between gauge
theories on non-commutative (NC) spacetime (non-commutative gauge
theories) and ordinary gauge theories 
in the background of constant NS-NS two-form field $B$
\cite{Connes:1998cr, Douglas:1998fm}. To be concrete, let us consider
a D3-brane in the background $B$-field in type IIB
string theory. When the $B$-field is polarized along the D3-brane,
then using T-duality and Fourier transformation the theory on the
D3-brane is shown to be equivalent with 4-dimensional $N\!=\!4$
supersymmetric gauge theory on the non-commutative spacetime defined
by
\begin{eqnarray}
  [x_i, x_j]=i\theta_{ij},
\end{eqnarray}
where the parameter $\theta$ specifies the extent of the
non-commutativity. Seiberg and Witten \cite{Seiberg:1999vs} showed
that these two (non-commutative and ordinary) descriptions are actually
stemmed from the different methods of regularization when derived
from string theory. According to them, the fields in each
description are related by some field redefinition, and the
actions in two descriptions are the same under this redefinition
(Seiberg-Witten transformation).
In this paper, we study small $\theta$ expansion. In order for
the calculation in both descriptions to be reliable, we choose the
region $\alpha' \gg \theta$. In this region, metrics in both
descriptions are almost flat, $\eta_{ij} + {\cal O}
\left((\theta/2\pi\alpha')^2\right)$. 
The small $\theta$ limit is equivalent with the small $B$ limit, 
\begin{eqnarray}
  2\pi\alpha' B_{ij} = -\frac{\theta_{ij}}{2\pi\alpha'}
+ {\cal O}\left(\left(\theta/2\pi\alpha'\right)^3\right).
\label{eq:bthe}
\end{eqnarray}
In this small $\theta$ and $B$ expansion, the effective actions in both
sides were shown to be the same \cite{Seiberg:1999vs}.

\begin{figure}[tdp]
\begin{center}
\leavevmode
\epsfxsize=50mm
\epsfbox{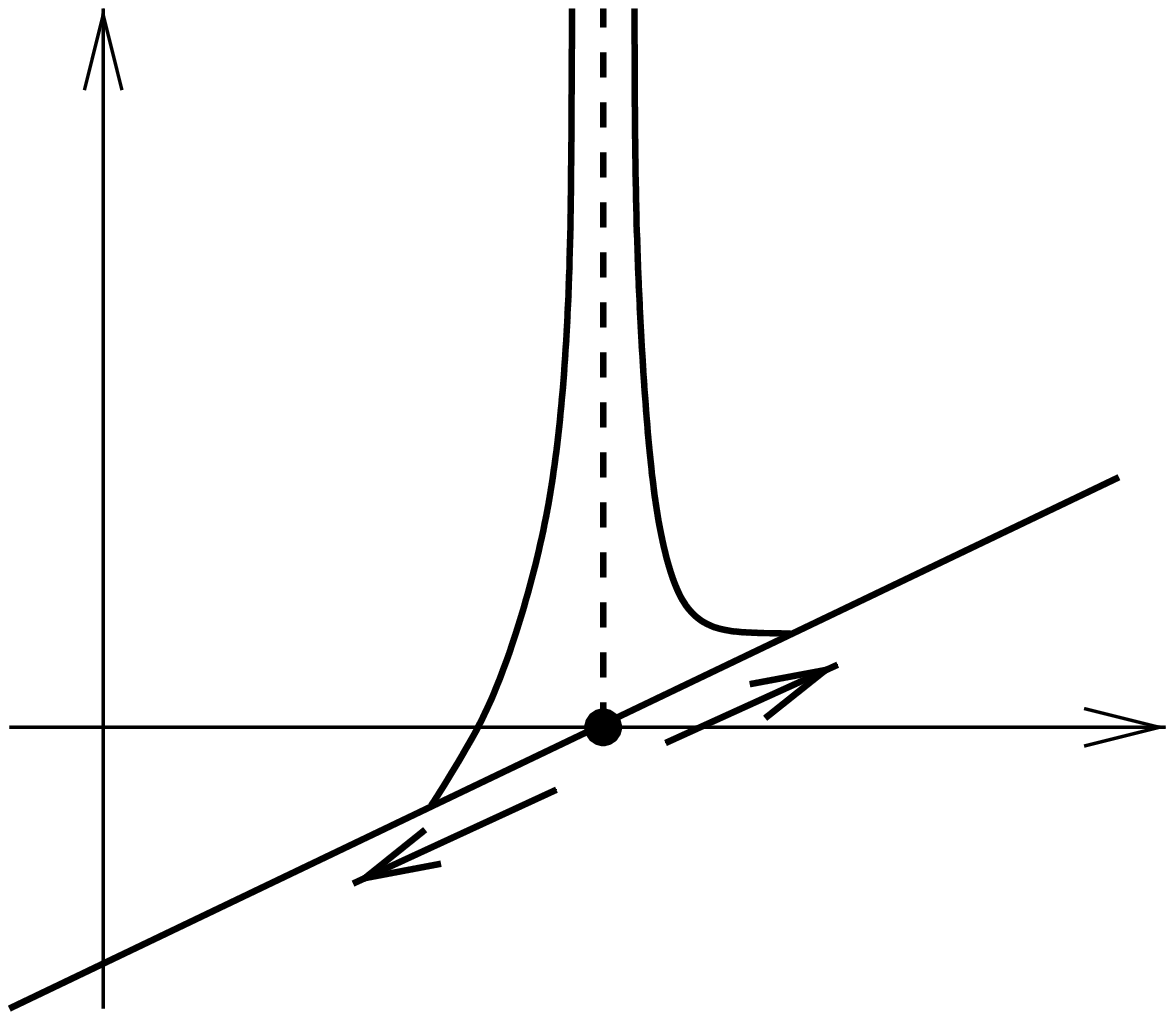}
\put(0,52){D3-brane}
\put(-120,110){D-string}
\caption{The D-string tilts against the D3-brane.}
\label{fig:tilt}
\end{center}
\end{figure}

So as to investigate theories on the non-commutative spacetime,
 the first step is to study the
properties of the solitons existing in those theories. Using the above
equivalence, monopoles and dyons in 4-dimensional non-commutative
gauge theory have been analyzed
\cite{Hashimoto:1999,Hashimoto:1999bc,Bak:1999,Hata:2000sj}. 
In ref.\ \cite{Hashimoto:1999}, using brane configurations in
the background $B$-field, the `non-commutative monopoles' were
analyzed through the brane interpretation of ref.\
\cite{Callan:1997}. The key observation of ref.\  
\cite{Hashimoto:1999} was that the stuck D-sting 
tilts in the $B$-field background. The existence of the
$B$-field is effectively the same as the existence of the magnetic
field on the D3-brane, and the magnetic force acting on the end of the
D-string is compensated by the tension of the tilted D-string
\cite{Hashimoto:1999bi} (see Fig.\ \ref{fig:tilt}).

It is very easy to see that the tilt of the D-string in the background
$B$-field is actually given by $\theta$. Let us consider a D3-brane in
this background. In the world volume $U(1)$ gauge theory on this
D3-brane, the usual BPS equation is
\begin{eqnarray}
  F_{ij} + B_{ij} = \epsilon_{ijk}\p_k \Phi,
\end{eqnarray}
where we turn on only a single scalar field $\Phi$. A point magnetic
charge preserving half of the supersymmetries of the theory is
described by the solution
\begin{eqnarray}
  \Phi = -\frac{g}{r} + \Half\epsilon_{ijk}B_{ij}x_k.
\label{phibsl}
\end{eqnarray}
This solution is depicted in fig.\ \ref{fig:tilt}.
The first singular term in the right hand side represents the stuck
D-string. The linear behavior of the second term indicates the tilt
of the D3-brane. The relative angle between the D3-brane and the
D-string is given by
\begin{eqnarray} 
2\pi\alpha'\Half\epsilon_{ijk}B_{ij} = 
-\Half\epsilon_{ijk}\theta_{ij}/2\pi\alpha',  
\label{eq:slope}
\end{eqnarray}
where we have introduced the parameter $2\pi\alpha'$ for defining the
dimensionless slope in the target space. In eq.\ 
(\ref{eq:slope}), we have adopted the limit $\theta \ll \alpha'$ and used
eq.\ (\ref{eq:bthe}). 

In this paper, we concentrate on monopoles in the non-commutative
$U(1)$ gauge theory. In the $U(1)$ case, there is a clear
understanding between the ordinary
and non-commutative gauge theories \cite{Seiberg:1999vs}, compared to
the non-Abelian case. It is possible to investigate the correspondence 
of the BPS equations in both sides. From the viewpoint of the brane
interpretation, monopoles are more suitable than instantons whose
non-commutative version were studied in refs. \cite{Seiberg:1999vs,
Marino:1999ni,Terashima:1999ui,Nekrasov:1998io,Braden:1999st,
Furuuchi:1999io}.

This paper is organized as follows. In sec.\ \ref{diracm}, we
solve the BPS equation in the non-commutative $U(1)$ gauge
theory. Then in sec.\ \ref{sec:brane}, we perform the Seiberg-Witten
transformation on the solution obtained in sec.\ \ref{diracm}, and
show that this exhibits an expected brane configuration of the tilted
D-string. In sec.\ \ref{sec:bpsor}, we analyze the relation
between the BPS equations in the non-commutative and ordinary
theories. In the commutative spacetime description, the non-linearly
realized supertransformation plays a crucial role. In 
sec.\ \ref{sec:rot}, we study the target space rotation which relates
the solution in sec.\ \ref{sec:brane} with the simple solution 
(\ref{phibsl}). Finally in sec.\ \ref{sec:conc}, we conclude with
future directions. In addition to the $U(1)$ case of our main
interest, the non-commutative $U(2)$ monopole and the non-commutative
1/4 BPS dyon are briefly studied in the commutative description in
sec.\ \ref{sec:u2} and Appendix A.

\section{Dirac monopole in non-commutative $U(1)$ gauge theory}
\label{diracm}

The soliton in the gauge theory is suitable for checking
correspondence between the Dirac-Born-Infeld (DBI) action \cite{dbi}
on the non-commutative spacetime and the DBI action with constant
NS-NS two-form background. We consider the simple situations, {\it i.e.},
the Dirac monopole and electrically charged particle (with source)
solutions in the $U(1)$ gauge theory. 

In this paper we concentrate on the effect of the
non-commutativity. Then the leading effect on non-commutativity to the
configurations is $\theta/r^2$. Therefore we do not take the DBI
theory but the Maxwell theory on the non-commutative spacetime. Before
we write the action, we comment on the justification we use the
Maxwell theory. In this approximation one may wonder higher derivative
generalization of the DBI action introduces the $\alpha'$ corrections
\begin{eqnarray}
  \frac{\alpha'}{r^2}, \hs{3ex}
  \frac{\alpha'^2}{r^4}, \hs{3ex}
  \frac{\alpha'\theta}{r^4}, \hs{3ex}
  \cdots ,
\end{eqnarray}
and the correction $\alpha'/r^2$ is larger than $\theta/r^2$.
However the corrections $(\alpha'/r^2)^k$ do not exist since by taking 
$\theta\rightarrow 0$ limit BPS solutions in the Maxwell theory are
also the ones in the DBI action \cite{Thorlacius:1997bi}. The correction
$\alpha'\theta/r^4$  may exist, but this is sub-leading
compared to $\theta/r^2$. Hence the $\theta/r^2$ effect is
accurately reproduced from the Maxwell theory.

The non-commutative $U(1)$ gauge theory with a Higgs field is
described by the following action,
\begin{eqnarray}
  S &=& \int d^4x~ \left( -\frac{1}{4}F_{\mu\nu} {*} F^{\mu\nu}
  +\frac{1}{2}D_{\mu}\Phi {*}D^{\mu}\Phi \right)~,
\end{eqnarray}
where we have defined the field strength and the covariant derivative
as
\begin{eqnarray}
  F_{\mu\nu} &=& \p_{\mu}A_{\nu} -\p_{\nu}A_{\mu}
  -i[A_\mu, A_\nu] ~,
  \\
  D_{\mu}\Phi &=&\p_\mu\Phi -i[A_{\mu},\Phi] ~.
\end{eqnarray}
We put the gauge coupling one for convenience. The commutator is
defined through the star product : $[A,B]\equiv
A{*}B-B{*}A$ and the star product is 
\begin{eqnarray}
  (f {*} g)(x) &\equiv& f(x)\exp (\frac{i}{2}\theta^{\mu\nu}
  \overleftarrow{\p_\mu}\overrightarrow{\p_\nu}) g(x) 
  = f(x)g(x) +\frac{1}{2} \{ f,g \}_P (x) +{\cal O}(\theta^2) ~,
\end{eqnarray}
where $\{ f,g \}_P (x)$ is the Poisson bracket,
\begin{eqnarray}
  \{ f,g \}_P (x) &=& i\theta^{\mu\nu}\p_\mu f(x) \p_\nu g(x) ~.
\end{eqnarray}

From the action the equations of motion for $A_\mu$ and $\Phi$ are
\begin{eqnarray}
  D^\mu F_{\mu\nu} &=& -i[\Phi,D_\nu\Phi] ~,
  \label{eomf}\\
  D^\mu D_\mu\Phi &=&0 ~,
\end{eqnarray}
and the Bianchi identity is
\begin{eqnarray}
  \epsilon^{\mu\nu\rho\sigma}D_\nu F_{\rho\sigma} &=& 0 ~.
  \label{bianchi}
\end{eqnarray}
The energy of this system is in the same form as for the ordinary
gauge theory except for changing the product by the star
product. Therefore the BPS equation for the static monopole is
\begin{eqnarray}
  \frac{1}{2}\epsilon_{ijk}F^{jk} &\equiv& B_i = D_i\Phi ~,
  \hs{3ex} i = 1, 2, 3 ~.
\end{eqnarray}
We calculate various quantities in the $\theta$ expansion and solve
the above equation to ${\cal O}(\theta)$ for studying the
non-commutative effect. The zero-th order solutions in $\theta$ are
\begin{eqnarray}
  A^{(0)}_1 &=& -\frac{g}{r(r+x_3)}x_2 ~,
  \\
  A^{(0)}_2 &=& \frac{g}{r(r+x_3)}x_1 ~,
  \\
  A^{(0)}_3 &=& 0 ~,
  \\
  \Phi^{(0)} &=& -\frac{g}{r} ~,
\end{eqnarray}
where the superscript means the order in
$\theta$. We take the solution with the Dirac string spreading on the
negative $x_3$ axis and the gauge is fixed by $A_0=\p^iA^{(0)}_i=0$.

By expanding the
equation (\ref{eomf}) to the first order in $\theta$ we obtain
\begin{eqnarray}
  \epsilon_{ijk}\p_j B^{(0)}_k +\epsilon_{ijk}\p_j B_k^{(1)} 
  -i\epsilon_{ijk}\{ A^{(0)}_j,B^{(0)}_k \}_P
  &=& -i\{ \Phi^{(0)}, \p_i\Phi^{(0)} \}_P ~.
\end{eqnarray}
Using $\p^iA_i^0 =0$ and $\p_i\Phi^0 =B_i^0$ we can easily solve this
equation for $B_a^1$ as
\begin{eqnarray}
  B_i^{(1)} &=& -i\{A_i^{(0)},\Phi^{(0)}\}_P +\p_i f ~,
\end{eqnarray}
with an arbitrary function $f$.
We substitute this solution into the Bianchi identity (\ref{bianchi})
and obtain the equation for $f$,
\begin{eqnarray}
  \Box f &=& 2i\p^i\{A_i^{(0)},\Phi^{(0)}\}_P ~.
\end{eqnarray}
The non-commutative effect appears as the form of the Poisson bracket,
to which $\theta^{0i}$ does not contribute. In the following we 
turn on only $\theta^{12}=\theta$. Then with the
boundary condition that the value of $f$ goes to zero asymptotically
we can solve $f$ as
\begin{eqnarray}
  f &=& -\theta g^2 \left( \frac{2x_3}{r^4} -\frac{1}{r^3} \right) ~.
\end{eqnarray}
In the same way we put $B_i^{(1)}$ into the BPS equation and obtain
the ${\cal O} (\theta)$ solution for $\Phi^{(1)}$ as
\begin{eqnarray}
  \Phi^{(1)} &=& \theta g^2 \left(\frac{1}{r^3} -\frac{2x_3}{r^4}
  \right) ~.
\end{eqnarray}
We summarize the BPS solutions in ${\cal O} (\theta)$
\begin{eqnarray}
  \Phi &=& -\frac{g}{r} +\theta g^2 \left(\frac{1}{r^3} 
    -\frac{2x_3}{r^4} \right) +{\cal O}(\theta^2)~,
  \label{solhiggs}\\
  B_i &=& \frac{gx_i}{r^3}  -i\{A_i^{(0)},\Phi^{(0)}\}_P +\p_i f 
  +{\cal O}(\theta^2) ~.
\end{eqnarray}
The $\theta/r^3$ term in the Higgs field is not proportional to
$\epsilon^{ijk}\theta_{ij}x_k (=\theta x_3)$ and is not invariant under
the Lorentz transformation that $\theta^{\mu\nu}$ is also properly
transformed. This seems strange. Since it is usually believed that the
(eigen) value of the Higgs field represents the brane configuration, it
should be invariant under the Lorentz transformation. To avoid this
problem the authors of \cite{Hashimoto:1999bc,Hata:2000sj} searched
the BPS solutions which have the Lorentz invariant eigenvalues of the
Higgs fields and discussed the brane tilting. In this Dirac monopole
case we can argue in the same way. Let us
take the zero-th order solution with the Dirac string spreading on the
positive $x_3$ axis. We calculate the Higgs field as
\begin{eqnarray}
  \Phi &=& -\frac{g}{r} +\theta g^2 \left(-\frac{1}{r^3} 
    -\frac{2x_3}{r^4} \right) +{\cal O}(\theta^2)~,
\end{eqnarray}
then we see only the part $\theta/r^3$ which is not Lorentz invariant
has an extra negative sign compared with the previous solution
(\ref{solhiggs}). Then if one wants a Lorentz invariant solution one
merges these solutions and obtain
\begin{eqnarray}
  \Phi &=& -\frac{g}{r} +\theta g^2 \left(
    -\frac{2x_3}{r^4} \right) +{\cal O}(\theta^2)~.
\end{eqnarray}
However there is another problem:
the tiling angle does not agree with the prediction from the brane
interpretation. As said in the introduction the D-string spreads from
the D3-brane with angle $\theta$. Therefore the configuration of the
Higgs field must respect this fact.  We can represent this fact by
the equation 
\begin{eqnarray}
  \Phi &=& -\frac{g}{|x_i+\frac{1}{2}\epsilon_{ijk}\theta^{jk}\Phi |}
  ~.
\end{eqnarray}
In the situation we consider, only $\theta^{12}=\theta$ is 
non-zero, the above equation says 
\begin{eqnarray}
  \Phi &=& -\frac{g}{r} -\theta x_3g^2\frac{1}{r^4} 
  +{\cal O}(\theta^2/r^4)
  ~.  \label{magare}
\end{eqnarray}
This is different from the Lorentz invariant part of the solution
(\ref{solhiggs}) by a factor 2.

This difference apparently appears when we consider the electrically
charged
particle in the NC $U(1)$ theory. The BPS equation for the
electrically charged
particle is
\begin{eqnarray}
  F_{i0} \;\;\;\equiv\;\;\; E_i &=& D_i\Phi ~,
\end{eqnarray}
and the zero-th order solutions are
\begin{eqnarray}
  A_0^{(0)} &=&\Phi^{(0)} = -\frac{g}{r} ~, \hs{3ex}
  \mbox{other fields} =0 ~.
\end{eqnarray}
As in the case of the Dirac monopole, we can easily solve the
equations of motion and the BPS condition in the NC theory 
and show the zero-th order solutions are also full order solution no
matter how we turn on the non-commutativity $\theta^{\mu\nu}$. 
From the brane interpretation when $B^{0i}\sim\theta^{0i}$ is
non-zero, the 
F-string is tilted with angle $B\sim\theta$. However we cannot see the 
informations of the tilt from the Higgs configuration.

This shows that the Higgs field in the non-commutative theory is not
a good object when compared with the brane interpretation. In the next
section we resolve this question.

\section{Brane interpretation}
\label{sec:brane}

\subsection{Seiberg-Witten transformation and brane interpretation}
\label{SWu1}

Callan and Maldacena revealed the BPS solution of the Higgs field
corresponds to the string structure \cite{Callan:1997}. The solution
solved in the previous section must be realized in the same
way. In ref.\cite{Hashimoto:1999bc, Bak:1999}, the NC $U(2)$ monopole
was discussed. The author of the 
ref. \cite{Bak:1999} discussed the Nahm equation in the NC gauge
theory and the effect of the non-commutativity in the Nahm equation
showed the D-strings slanted with slope $\theta$. In
ref. \cite{Hashimoto:1999bc}, eigenvalues of the Higgs field were
investigated in the NC gauge theory and their brane interpretations
were investigated. The eigenvalue equation of a matrix valued function
$M$ in the non-commutative space takes the form
\begin{eqnarray}
  M {*} \vec{v} &=& \lambda {*} \vec{v} ~,
  \label{eigen}
\end{eqnarray}
where $\vec{v}$ and $\lambda$ are the eigenvector and its eigenvalue,
respectively. In this form the eigenvalue 
is the same as the expected form, {\it i.e.}, D-string is tilted. However
$\lambda$ in (\ref{eigen}) is not gauge invariant and we have not
known other forms taking informations on the brane configurations in
NC gauge theory more properly.  

We have argued in the previous section that the configurations in the
NC side does not match the brane interpretation. In the
NC theory, since the coordinates do not commute with each
other, functions written by only the coordinates are also
operators. However we do not know the appropriate method for extracting
the gauge singlet c-number quantities\footnote{The Higgs field in the
  NC $U(1)$ theory is not singlet.}. 
Moreover the tilted brane is expected in the commutative spacetime,
not in the non-commutative spacetime. Therefore we
insist that it is appropriate to study the brane interpretation in the
ordinary gauge theory which is equivalent with the NC gauge theory. 

Seiberg and Witten \cite{Seiberg:1999vs} showed that non-commutative and
ordinary gauge theories are equivalent under the following relation of
the gauge fields
\begin{eqnarray}
  \what{A}_\mu &=& A_\mu -\frac{\theta^{\rho\delta}}{4}
  \left\{ A_\rho, \p_\delta A_\mu +F_{\delta \mu} \right\} 
  +{\cal O}(\theta^2) ~,
  \label{gausw}\\
  \what{\Phi} &=& \Phi -\frac{\theta^{\rho\delta}}{4} 
  \{A_\rho, \p_\delta\Phi +D_\delta \Phi \} +{\cal O}(\theta^2) ~,
  \label{eq:phisw}\\
  \what{F}_{\mu\nu} &=& F_{\mu\nu} +\frac{\theta^{\rho\delta}}{4}
  \left( 2\left\{F_{\mu\rho}, F_{\nu\delta}\right\} 
    -\left\{A_\rho, D_\delta F_{\mu\nu} +\p_\delta F_{\mu\nu}
    \right\}\right) +{\cal O}(\theta^2) ~,
\end{eqnarray}
where $\{ A,B\}= AB +BA$ is the anti-commutator. These relations are
obtained by requiring
\begin{eqnarray}
  \what{A}(A) +\what{\delta}_{\what{\lambda}}\what{A}(A)
  &=& \what{A}(A+\delta_\lambda A) ~,
\end{eqnarray}
with infinitesimal $\lambda$ and $\what{\lambda}$. We denote $\what{A}$
as the gauge field in the non-commutative side and $A$ is the one 
in the ordinary gauge theory. Using these mappings, we can easily
obtain
the configurations for the non-commutative Dirac monopole in the
ordinary gauge theory, 
\begin{eqnarray}
  \Phi &=& -\frac{g}{r} +\theta g^2 ( -\frac{x_3}{r^4})
   +{\cal O}(\theta^2)
  ~,\label{orphi}\\
  B_1 &=& g\frac{x_1}{r^3} \left( 1 +4\theta x_3 \frac{g}{r^3} \right)
   +{\cal O}(\theta^2)
  ~,\\
  B_2 &=& g\frac{x_2}{r^3} \left( 1 +4\theta x_3 \frac{g}{r^3} \right)
   +{\cal O}(\theta^2)
  ~,\\
  B_3 &=& g\frac{x_3}{r^3} \left( 1 +4\theta x_3 \frac{g}{r^3} \right)
  -2\theta g^2\frac{1}{r^4}  +{\cal O}(\theta^2)~.
  \label{orb}
\end{eqnarray}
Then the Higgs field is invariant under the Lorentz transformation and
the same as (\ref{magare}). The problem that the Higgs field is not
invariant under the Lorentz transformation which occurs in the
non-commutative side disappears and the bending angle from D3-brane
exactly matches with the expected one.

The above discussion also holds for the non-commutative electrically
charged particle. The corresponding solution in the ordinary theory is
easily obtained from the mappings as
\begin{eqnarray}
  \Phi &=& -\frac{g}{r} -g^2\frac{\theta^{0i}x_i}{r^4}  
  +{\cal O}(\theta^2)~,
\end{eqnarray}
which takes the expected form. In the NC gauge theory, the electrically
charged particle does not receive the non-commutative effect. On the
other hand, in the ordinary theory, the $B$-field coupling in the DBI
action generates corrections to the configuration of the electrically
charged particle.

In the DBI action the $B$-field always appears in the combination
$F^{\mu\nu}+B_{NS}^{\mu\nu}$, and so $B^{ij}_{NS}$ behaves as the
magnetic field and $B_{NS}^{0i}$ behaves as the electric field. We
recognize this behavior from the forms of the solutions.
The ``electric field'' $B_{NS}^{0i}$ changes the configuration of
the electrically charged particle and the ``magnetic field''
$B_{NS}^{ij}$ alters that of the Dirac monopole.

So far we have concentrated on the solutions of the Higgs field. Later 
we reanalyze the DBI action for small $B$-field
and confirm that the configurations considered in this section are the
BPS solutions in the ordinary theory. 

A comment is in order: the $\theta$ expansion is well defined for
$\theta \ll r^2$ since the dimensionless parameter for the $\theta$
expansion is $\theta g/r^2$. In that region the value of the Higgs
field is reliable and the D-string slants with angle $\theta$ (the
equation (\ref{magare}) is reliable). Therefore we naturally regard
the Dirac monopole in the NC theory as the D-string attached to
D3-brane with uniform magnetic fields. We discuss this relation in
section \ref{sec:rot}.


\subsection{Non-commutative $U(2)$ monopole and Seiberg-Witten
  transformation}
\label{sec:u2}

As seen in sec.\ \ref{diracm}, for the $U(1)$ BPS Dirac monopole, the
solution in the non-commutative side (named {\bf (I)}) does not
exhibit the appropriately slanted D-string. The configuration after
the Seiberg-Witten map (\ref{eq:phisw}) gives the precise tilt of the
D-string. Therefore, although the non-commutative $U(2)$ monopole was
already considered in ref.\ \cite{Hashimoto:1999bc}, it is natural to
study their commutative counterparts.

The monopole solution in the non-commutative super Yang-Mills theory
obtained in ref.\ \cite{Hashimoto:1999bc} is
\begin{eqnarray}
&&  \what{A}_i = \epsilon_{aij} \frac{x_j}{r}W(r)\Half\sigma_a
+\theta_{ij}x_j\frac{1}{4r^2}W(r)\biggl(W(r) + 2F(r)\biggr) 
\Half\unit+{\cal O}(\theta^2),\\
&& \what{\Phi} = \frac{x_a}{r}F(r)\Half\sigma_a +{\cal O}(\theta^2), 
\end{eqnarray}
where we have defined 
\begin{eqnarray}
  F(r) \equiv C \coth (Cr) -\frac{1}{r}, \qquad
  W(r) \equiv \frac1r - \frac{C}{\sinh (Cr)} ,
\end{eqnarray}
with a dimension-ful parameter $C$. Note that there is no 
${\cal O}(\theta^1)$ correction $\Phi^{(1)}$ in the Higgs field.
Performing the Seiberg-Witten transformation (\ref{eq:phisw}), we
obtain the configuration for $\Phi$ in the commutative description
(named {\bf (II)}) as
\begin{eqnarray}
  \Phi = \frac{x_a}{r}F(r)\Half\sigma_a 
+\epsilon_{ijk}\frac{x_k}{r^2}
W(r) F(r) \biggl(
2-r W(r)
\biggr)\Half\unit+{\cal O}(\theta^2).
\label{eq:u2mono}
\end{eqnarray}
The eigenvalues of this matrix $\Phi$ are of course gauge invariant.
Near the infinity of the worldvolume we have the asymptotic expansion 
for the eigenvalues  as
\begin{eqnarray}
  \lambda = \pm
  \left(
    C-\frac1{r}
  \right)\Half 
- \frac{\epsilon_{ijk}x_i\theta_{jk}}{8r^3}
  \left(
    C-\frac1{r}
  \right)
+{\cal O}(\theta^2).
\end{eqnarray}
Remarkably, this asymptotic expression is the same as the one obtained
in ref.\ \cite{Hashimoto:1999bc} where the ${\cal O}(\theta)$
eigenvalues are generated using the `non-commutative eigenvalue
equation'. Since in ref.\ \cite{Hashimoto:1999bc} 
this expression was shown to match the tilted D-string configuration
with the proper slope $\theta$, we see that the Seiberg-Witten
transformed configuration in {\bf (II)} exhibits the correct
configuration of the slanted D-string. 

\begin{figure}[htdp]
\begin{center}
\leavevmode
\epsfxsize=100mm
\epsfbox{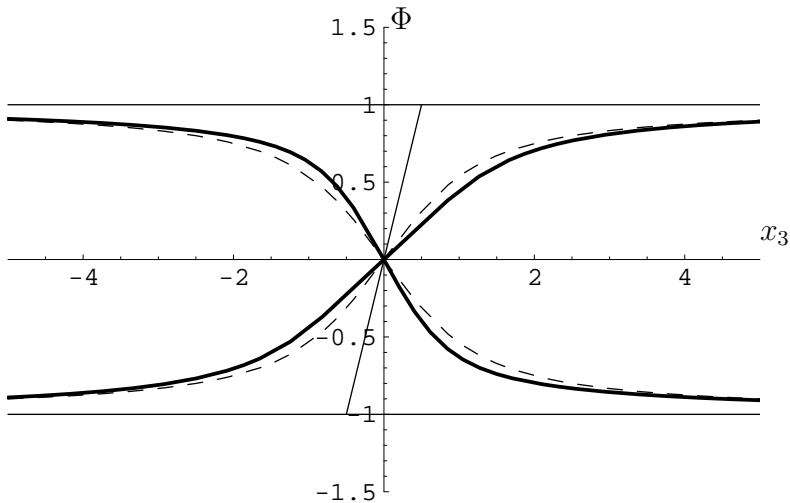}
\put(0,100){$x_3$}
\put(-140,180){$\Phi$}
\caption{The eigenvalues in the commutative description of the
  non-commutative $U(2)$ monopole. The configuration written by the
  bold curve is regular everywhere. The dashed curves denote the
  D3-brane configuration with $\theta =0$. The slanted line indicates
  the D-string. Here we have put $C=2$ and $\theta=1/2$ for
  simplicity.}  
\label{fig:u2}
\end{center}
\end{figure}

In addition, the configuration (\ref{eq:u2mono}) has another nice
property: the configuration is regular even at the origin
$r=0$. (The eigenvalues obtained in ref.\ \cite{Hashimoto:1999bc} were 
singular at the origin.) Since we can prove the following relation 
\begin{eqnarray}
  \lambda = \pm \Half F
\left(|x_i - \Half\epsilon_{ijk}\theta_{jk}\lambda|\right) ,
\end{eqnarray}
up to ${\cal O}(\theta^2)$, we understand that the D-string is
suspended really along the line  
\begin{eqnarray}
  x_i = \Half \epsilon_{ijk}\theta_{jk}\lambda ,
\end{eqnarray}
and has a tilt $\theta$. The interesting is that the tilted
D-string can be read not only from the asymptotic region but also from 
everywhere. The D3-brane configuration
in the commutative description {\bf (II)} (the eigenvalues of eq.\
(\ref{eq:u2mono})) is depicted in fig.\ \ref{fig:u2}. 

In a similar manner, we obtain the magnetic field in {\bf (II)} using
the Seiberg-Witten transformation. The result for the gauge field is
\begin{eqnarray}
  A_i =
\widehat{A}^{(0)}_i + \theta_{ij}x_j \frac{1}{2r^2} 
W(r) \left(W(r) + F(r) + r \frac{\p W(r)}{\p r}\right) \Half\unit
+{\cal O}(\theta^2),
\label{eq:magu2}
\end{eqnarray}
and it is straightforward to obtain the magnetic field from this
expression. Expanding near the infinity, this also coincides with the
result obtained in ref.\ \cite{Hashimoto:1999bc}\footnote{Note that
  the expression of the magnetic field written in ref.\
  \cite{Hashimoto:1999bc} contains a typo of a factor 2.}.
The magnetic field configuration from eq.\ (\ref{eq:magu2}) is also
regular at the origin.

In Appndix A, we apply this procedure also for the non-commutative
$U(3)$ monopole and the non-commutative $1/4$ BPS dyon which were
studied in ref.\ \cite{Hata:2000sj}.

\section{BPS condition for ordinary gauge theory}
\label{sec:bpsor}

In the previous section we have concentrated on the configurations of
the field in the ordinary gauge theory and do not paid attention to
the action. Seiberg and Witten showed that the DBI action for the
small $B$-field is equal to the non-commutative DBI action for small
$\theta$ with $\alpha'$ fixed, {\it i.e.} $\theta/\alpha' \ll 1$,  by the
redefinition of fields and couplings. From this equivalence, we can
consider the BPS equation for the ordinary gauge theory and check
whether the
configuration (\ref{orphi}) $\cdots$ (\ref{orb}) satisfies the BPS
equation or not.

We use the DBI action for the ordinary gauge theory with a scalar
field, 
\begin{eqnarray}
  {\cal L} &=& \sqrt{\det \Big( -g_{\mu\nu}
    +2\pi\alpha' (B_{\mu\nu}+F_{\mu\nu})
    +(2\pi\alpha')^2 \p_{\mu}\Phi\p_{\nu}\Phi
    \Big)} ~,
  \label{orlag}
\end{eqnarray}
where the numerical factor is omitted. Now we consider the situation
$B_{\mu\nu}=-\theta_{\mu\nu}/(2\pi\alpha')^2$ and the metric
$g_{\mu\nu}$ is flat. We do not expand the lagrangian (\ref{orlag})
under the small $B$-field, because the obtained action has many
interactions and is not suitable for picking up physical
meanings. Therefore we consider the DBI action itself. 

In the work of Seiberg and Witten, the equivalence between the 
non-commutative and ordinary gauge theories was shown in the
approximation of slowly varying fields. However the solutions of the 
Dirac monopole in the NC gauge theory do not vary slowly, we shall
investigate whether (\ref{orphi}) $\cdots$ (\ref{orb}) are the BPS
solution in the ordinary theory. For this purpose let us study the BPS
condition and see the solutions satisfy the BPS equation.

For discussing the BPS condition, it is more useful to consider the
supersymmetrized theory and the condition for some supersymmetry
remained unbroken than to study the minimal bound of the energy for the
system. In the case of monopole we must consider the $N=2$
supersymmetric DBI action in 4 dimensions and study the supersymmetric
transformation for fermions. If $B$-field is zero, the linearly
realized supertransformations for fermions are (in 6 dimensional
notation),
\begin{eqnarray}
  \delta \lambda &=& F_{MN}\Sigma^{MN}\eta, \hs{3ex}
  M,N =0,\cdots 5 ~,
\end{eqnarray}
where $A_M$ for $M=0,\cdots, 3$ are gauge fields in four dimensional
spacetime and $A_4=\Phi$ and $A_5$ is another scalar field. Then 
the BPS condition for the (Dirac) monopole becomes simple;
$F_{MN}\Sigma^{MN}$ has some zero eigenvalues. On the other hand, if
$B$ is nonzero, all the linearly realized supertransformations are
broken, and the unbroken supertransformations are some combinations of 
the linearly and non-linearly realized ones. Thus we must see the
non-linearly realized supertransformations. The $N=2$ DBI action is
obtained as the model of partial breaking of $N=4$ supersymmetry down to 
$N=2$ \cite{ketov, tsey} and we can read the non-linear
transformations for the broken
supersymmetries. However fortunately the nonzero fields are only one
Higgs scalar and the space components of the gauge fields, and they are
static. Then we need not to know the full non-linear ones but we only
see the $N=1$ part which generate shifts for fermions as
\begin{eqnarray}
  \delta \lambda_+ &=& (F^+_{mn}\! +\!B^+_{mn})
  \sigma^{mn} \eta_+ , 
  \label{n2li}\\
  \delta \lambda_- &=& (F^-_{mn}\! +\!B^-_{mn})
  \sigma^{mn} \eta_- , 
  \label{n2li2}\\
  \what{\delta} \lambda_+ &=& \frac{1}{4\pi\alpha'}\left(
    1 -(2\pi\alpha')^2{\rm Pf} (F_{mn}\!+\!B_{mn}) 
    +\sqrt{\det (\delta_{mn}+2\pi\alpha'(F_{mn}\!+\!B_{mn}))}
  \right)\chi_+ ~,
  \label{n2nonl}\\
  \what{\delta} \lambda_- &=& \frac{1}{4\pi\alpha'}\left(
    1 +(2\pi\alpha')^2{\rm Pf} (F_{mn}\!+\!B_{mn}) 
    +\sqrt{\det (\delta_{mn}+2\pi\alpha'(F_{mn}\!+\!B_{mn}))}
  \right)\chi_- ~,
  \label{n2nonl2}\\
  &&  m,n =1,2,3,4, \nonumber
\end{eqnarray}
where (\ref{n2li}) and (\ref{n2li2}) are linear ones and
(\ref{n2nonl}) and (\ref{n2nonl2}) are non-linear ones\footnote{We use
  an unusual decomposition. We decompose a Weyl fermion in 6
  dimensions into $SO(4)=SU(2)_+\times SU(2)_-$ fermions $\lambda_+$
  ({\bf 2},{\bf 1}) and $\lambda_-$ ({\bf 1},{\bf 2}). 
  $F^+$ is the tensor transforming as ({\bf 3},{\bf 1}) and
  $F^-$ is the tensor transforming as ({\bf 1},{\bf 3}). Notice that
  this $SO(4)$ is not the Lorentz group in our spacetime ($M=0,1,2,3$)
  but the rotation on the plane $M=1,2,3,4$.
}, and we have defined $4{\rm Pf} F =\tr (F^+)^2-\tr (F^-)^2$.
In this expression we have already put unnecessary fields to zero.
We consider the $B$-field is non-zero for the space direction, namely
$B^{12}=\theta/\alpha'^2$, since $B^{0i}$ does not affect the monopole
configuration to the first order in $B$, and we take the metric
flat. Notice that $m=4$ is not the spacetime direction but
$F_{m4}=\p_m\Phi$ and $B_{m4}=0$. The above transformation is the same
as the $N=1$ linear and non-linear transformations \cite{bagger} in
the Euclidean 4 dimensional space except for replacing the fourth gauge
field by the Higgs field $\Phi$ and putting $\p_4 =0$. This fact
is natural, since from the 6 dimensional aspect to set $\p_0=0$ which
means static, $\p_5=0$, $A_0=A_5=0$ and half of fermion to zero, the
theory reduces to the Euclidean 4 dimensional $N=1$ supersymmetric gauge
theory and the linearly and non-linearly realized supertransformations
also reduce to the $N=1$ ones.

Now we have tools for studying the 1/2 BPS condition for the monopole
with non-zero $B$. In the situation we now consider $B$ and $F$ have
the following forms,
\begin{eqnarray}
  B_{mn} &=&\left(
    \begin{array}{cccc}
      0&B\\
      -B&0\\
      &&0\\
      &&&0
    \end{array}
  \right)~,
  \hs{3ex}
  B=-\theta/4\pi^2\alpha'^2 ~,
  \label{bnokatati}\\
  F_{mn}&=&\left(
    \begin{array}{cccc}
      0&B_3 &-B_2&-\p_1\Phi\\
      -B_3&0&B_1&-\p_2\Phi\\
      B_2&-B_1&0&-\p_3\Phi\\
      \p_1\Phi&\p_2\Phi&\p_3\Phi&0
    \end{array}
    \right)~,
    \hs{3ex}
    B_i =\frac{1}{2}\epsilon_{ijk}F^{jk}~,
    (i=1,2,3),
    \label{fnokatati}
\end{eqnarray}
and then ${\rm Pf} B=0$ obeys. $B^+$ and $F^+$ are defined as
\begin{eqnarray}
  B^+_{mn} &=& \frac{1}{2}( B_{mn} +\widetilde{B}_{mn} ), 
  \hs{3ex}
  \widetilde{B}_{mn} =\frac{1}{2}\epsilon_{mnpq}B^{pq}
  ~,\\
  F^+_{mn} &=& \frac{1}{2}( F_{mn} +\widetilde{F}_{mn}),
  \hs{3ex}
  \widetilde{F}_{mn} =\frac{1}{2}\epsilon_{mnpq}F^{pq}
  ~.
\end{eqnarray}
Since $F_{mn}$ goes to zero asymptotically, the combination
$\delta (\eta_+) +\what{\delta}(\chi'_+)$,
\begin{eqnarray}
  \chi'_+ &=& -\frac{4\pi\alpha'}
  {1 -(2\pi\alpha')^2{\rm Pf}B_{mn} 
    +\sqrt{\det (\delta_{mn}+2\pi\alpha' B_{mn})}}
  B^+_{mn}\sigma^{mn}\eta_+
  \\
  &=& -\frac{4\pi\alpha'}
  {\left( 1+\sqrt{1 +(2\pi\alpha')^2 B^2}\right)}
  B^+_{mn}\sigma^{mn}\eta_+~,
  \label{ahhhh}
\end{eqnarray}
is the unbroken supertransformation. Other supertransformations are
all broken and then $N=1$ supersymmetry is unbroken\footnote{When we
  consider anti monopole $B_i=-\p_i\Phi$, we must consider the
  combination of $\eta_-$ and $\chi_-$ for the unbroken
  supertransformations. 
}.
Therefore the unbroken supertransformation must be $\delta (\eta_+)
+\what{\delta}(\chi'_+)$ ($\chi'_+$ does not change, {\it i.e.}
eq. (\ref{ahhhh}))  everywhere and the BPS condition is,
\begin{eqnarray}
  (\delta (\eta_+) +\what{\delta}(\chi'_+)) \lambda_+ &=& 0 ~.
\end{eqnarray}
This is written explicitly as
\begin{eqnarray}
  0&=& (F^+_{mn} +B^+_{mn})\\
  &&
  -\frac{1 -(2\pi\alpha')^2{\rm Pf} (F_{mn}+B_{mn})
    +\sqrt{\det (\delta_{mn} +2\pi\alpha'(F_{mn}+B_{mn}))}}
  {1+\sqrt{1 +(2\pi\alpha')^2B^2}}B^+_{mn} ~.
\end{eqnarray}
We expand the right hand side to the first order in $B$ and second
order in $F$ since we consider the linearized Maxwell theory. This
approximation is equivalent to that we consider the Maxwell theory in
the NC gauge theory and study the first order in $\theta$. Then we
obtain the BPS condition,
\begin{eqnarray}
  F_{mn}^+ +B_{mn}^+\frac{(2\pi\alpha')^2}{8}
  (\tr F\widetilde{F}-\tr F^2)   &=&0  ~.
\end{eqnarray}
When we substitute (\ref{bnokatati}) and (\ref{fnokatati}) into
this condition, we finally obtain the following condition to the first
order in $\theta$,
\begin{eqnarray}
  B_1 &=& \p_1\Phi ,\\
  B_2 &=& \p_2\Phi ,\\
  B_3 &=& \p_3\Phi -\theta g^2\frac{1}{r^4} .
\end{eqnarray}
We can easily show the Dirac monopole solution in the ordinary theory
(\ref{orphi}) $\cdots$ (\ref{orb}) satisfies these equations. In the
end we have shown that, using the mappings (\ref{gausw}) and
(\ref{eq:phisw}), the BPS equation in the NC theory is transformed to
the BPS equation in the ordinary theory.

\section{Target space rotation}
\label{sec:rot}

\subsection{Reproduction of the solution} 
\label{sec:rotation}

The brane configuration obtained in sec.\ \ref{diracm} in the
ordinary gauge theory, (\ref{orphi}), has the desired property that
the D-string is slanted 
with the slope $\theta$. Now, a natural question arises --- how this
solution in {\bf (II)} is related to the
configuration (\ref{phibsl}) (named {\bf (III)}) considered in the
introduction? The difference between the two originates in only the
way of putting the coordinate system in the target space: they are
related by the target space rotation by the angle defined by $\theta$
(\ref{eq:slope}). 

\begin{figure}[bp]
\begin{center}
\leavevmode
\epsfxsize=100mm
\epsfbox{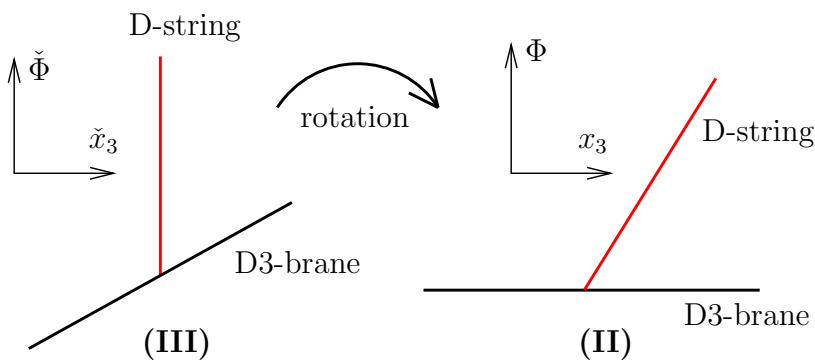}
\put(-70,77){$x_3$}
\put(-255,77){$\check{x}_3$}
\put(-30,10){D3-brane}
\put(-175,85){rotation}
\put(-200,30){D3-brane}
\put(-278,103){$\check{\Phi}$}
\put(-90,110){$\Phi$}
\put(-240,120){D-string}
\put(-23,80){D-string}
\put(-70,0){{\bf (II)}}
\put(-235,0){{\bf (III)}}
\caption{Rotation in the target space from {\bf (III)} to {\bf (II)}. }
\label{fig:rot}
\end{center}
\end{figure}

The BPS equation of the ordinary gauge theory adopted in the
introduction is 
\begin{eqnarray}
  \check{F}_{ij} + B_{ij} = \epsilon_{ijk}\check{\p}_k 
\check{\Phi},
\label{eq:BPSli}
\end{eqnarray}
and using (\ref{eq:bthe}) its solution is
\begin{eqnarray}
  \check{\Phi} = -\frac{g}{\check{r}} 
- \frac{1}{(2\pi\alpha')^2} \theta_{12} \check{x}_3,
\qquad
\Half \epsilon_{ijk} \check{F}_{jk} =
\frac{g \check{x}_i}{\check{r}^3},
\label{eq:magtil}
\end{eqnarray}
where the check indicates that the variables are in the description
{\bf (III)}. We have turned on only the
$\theta_{12}$ component, and therefore the configuration is slanted in
the direction along $\check{x}_3$. The target space rotation which
may relate {\bf (II)} and {\bf (III)} is 
\begin{eqnarray}
  \left(
    \begin{array}{c}
      2\pi\alpha' \Phi \\
      x_3
    \end{array}
  \right)
=
  \left(
    \begin{array}{cc}
      \cos \varphi & -\sin\varphi \\
      \sin \varphi & \cos\varphi 
    \end{array}
  \right)
  \left(
    \begin{array}{c}
      2\pi\alpha' \check{\Phi} \\
      \check{x}_3
    \end{array}
  \right),
\label{eq:rotation}
\end{eqnarray}
while the other coordinates are left invariant
($\check{x}_1 = x_1, \check{x}_2 =x_2$). Note that, as
mentioned in the introduction, we must multiply the factor
$2\pi\alpha'$ on the scalar field so as to adjust the dimensions. 
It is easy to see that the rotation angle $\varphi$ should be given by
$\varphi =-\theta_{12}/2\pi\alpha' 
+ {\cal O}\left((\theta/2\pi\alpha')^3\right)$. 
By substituting the solution (\ref{eq:magtil}) in {\bf (III)} into
eq.\ (\ref{eq:rotation}), we have  
\begin{eqnarray}
  \Phi & = & -\frac{g}{\check{r}}+ {\cal O}(\theta^2),
\label{eq:phie}\\
 x_3 & = & \theta_{12} \frac{g}{\check{r}} + \check{x}_3+ {\cal
  O}(\theta^2). 
\label{eq:x3}
\end{eqnarray}
We have chosen the value of $\varphi$ so that $\Phi$ vanishes
asymptotically. From eq.\ (\ref{eq:x3}) we have a relation
\begin{eqnarray}
  \check{r} = r
  \left(
    1-\theta_{12}\frac{g x_3}{r^3}
  \right)+ {\cal O}(\theta^2).
\end{eqnarray}
Therefore combining this with eq.\ (\ref{eq:phie}), finally we obtain 
\begin{eqnarray}
  \Phi = -\frac{g}{r} -\theta_{12}\frac{g^2x_3}{r^4}+
 {\cal O}(\theta^2),
\end{eqnarray}
which coincides with the solution in the previous section
(\ref{orphi}). 

From the very naive argument presented in this subsection, we have
seen that the solution in {\bf (II)} is easily obtained through the
target space rotation from {\bf (III)}. Since we have seen in 
sec. \ref{sec:brane} that the solution in the NC theory {\bf (I)} is
related to  {\bf (II)} through the Seiberg-Witten map, hence we
have three equivalent descriptions.

\subsection{Reproduction of the BPS equation}
\label{sec:repro}

We have seen that the BPS equation in the description {\bf (I)}
corresponds to the unusual BPS condition which preserves a certain
combination of linearly and non-linearly realized supersymmetries in
{\bf (II)}. Now, as seen in sec.\ \ref{sec:rotation}, the solution of
this unusual BPS equation is obtained by the target space rotation
(\ref{eq:rotation}) from the solution in the description of {\bf
  (III)}. In {\bf (III)} where the D3-brane is slanted, it is enough
to consider linearly realized supersymmetries, and the story becomes
considerably 
simple. Thus it might be natural to study how the rotation acts on the
BPS equation. In this subsection, we shall see that the BPS conditions
in {\bf (II)} and {\bf (III)} are related with each other under the
rotation. 

The BPS equation in {\bf (II)} reads
\begin{eqnarray}
  B_i + \delta_{i3}\theta_{12}g^2 \frac{1}{r^4} = \p_i \Phi.
\label{eq:BPScom}
\end{eqnarray}
We want to derive this equation from the BPS equation (\ref{eq:BPSli})
in {\bf (III)} by the rotation (\ref{eq:rotation}). First, from the
relation 
\begin{eqnarray}
  \check{\Phi} (\check{x}) = 
\Phi(x) -\frac{\theta_{12}}{(2\pi\alpha')^2} x_3,
\end{eqnarray}
using eq. (\ref{eq:rotation}), we have
\begin{eqnarray}
  \check{\Phi} (\check{x}) = 
\Phi(x_1, x_2, \check{x}_3-\theta_{12}\check{\Phi})
-\frac{\theta_{12}}{(2\pi\alpha')^2} \check{x}_3.
\end{eqnarray}
Thus the derivative with respect to $\check{x}_3$ is 
\begin{eqnarray}
  \check{\p}_3 \check{\Phi}(\check{x})
&&=
\frac{\p(\check{x}_3-\theta_{12}\check{\Phi})}{\p\check{x}_3}
\p_3 \Phi(x)
-\frac{\theta_{12}}{(2\pi\alpha')^2}\nn\\
&&=
\left(
  1-\theta_{12}\frac{\p\check{\Phi}}{\p\check{x}_3}
\right)
\p_3 \Phi(x)
-\frac{\theta_{12}}{(2\pi\alpha')^2}.
\end{eqnarray}
Therefore for the right hand side of eq.\ (\ref{eq:BPSli}) we have
\begin{eqnarray}
  \check{\p}_i \check{\Phi} 
=
\p_i \Phi(x)
-\delta_{3i} \frac{\theta_{12}}{(2\pi\alpha')^2} -\theta_{12}\p_3\Phi
\p_i \Phi.
\label{eq:pphi}
\end{eqnarray}
On the other hand, in the left hand side of eq.\ (\ref{eq:BPSli}), the
term containing the $B$-field is changed to
\begin{eqnarray}
  \Half\epsilon_{ijk}B_{jk} = -\frac{1}{(2\pi\alpha')^2}
\delta_{3i}\theta_{12},
\end{eqnarray}
hence this cancels the constant term in eq.\ (\ref{eq:pphi}) in the
right hand side of eq.\ (\ref{eq:BPSli}).
Now the magnetic field is expanded as
\begin{eqnarray}
  \check{B}(\check{x}) 
= \check{B}(x_1, x_2, x_3 + \theta_{12}\Phi) 
=  \check{B}(x)  + \theta_{12}\Phi\p_3 \check{B}(x) +
{\cal O}(\theta^2). 
\end{eqnarray}
Note that, as seen in eq.\ (\ref{eq:magtil}), the magnetic field
function $\check{B}$ is equal to the 
zero-th order solution $B^{(0)}$. Thus we can rewrite the BPS equation
(\ref{eq:BPSli}) as 
\begin{eqnarray}
B^{(0)}(x) + \theta_{12}\Phi^{(0)}\p_3B^{(0)}(x) 
=\p_i \Phi -\theta_{12}\p_3 \Phi^{(0)} \p_i \Phi^{(0)}.
\end{eqnarray}
This is the same as the BPS equation (\ref{eq:BPScom}) in {\bf (II)},
using the explicit solution for $B$ in {\bf (II)}.

\section{Conclusion and discussion}
\label{sec:conc}

In this paper, we have analyzed the non-commutative BPS Dirac monopole
in the three different descriptions: 
{\bf (I)} the solution of the BPS equation in the non-commutative
$U(1)$ gauge theory,  
{\bf (II)} the solution of the BPS equation which preserves a 
combination of the linearly and non-linearly realized supersymmetries 
in the ordinary $U(1)$ gauge theory in the $B$-field background, and
{\bf (III)} the solution of the usual BPS equation from linearly
realized supersymmetries in the ordinary $U(1)$ gauge theory in the
constant magnetic field. 
For small $\theta$ and small $B$ approximation, we have shown that 
these three descriptions are related with each other as follows: {\bf
  (I)} and {\bf (II)} are related by the Seiberg-Witten
transformation, and {\bf (II)} and {\bf (III)} are by the target space 
rotation. We have confirmed that the non-commutative Dirac monopole
matches perfectly with the tilted D-string configuration in the
$B$-field background.

The solution for the scalar field $\widehat{\Phi}$ obtained in {\bf
  (I)} is not invariant under the simultaneous rotation of $x_i$ and 
$\theta_{ij}$. However, performing the Seiberg-Witten transformation
on this solution, then we obtain a rotation-invariant solution for
$\Phi$ in the commutative description {\bf (II)}. This solution
exhibits precisely the configuration of the D-string tilted due to the 
existence of the $B$-field. This solution is also obtained by
the target space rotation from the simple solution in 
{\bf (III)}. This target space rotation is concerning the plane
spanned by the scalar field and  the worldvolume coordinate along the
non-commutativity.

Furthermore, also the BPS equations for each description have been
shown to be related with each other by the above  
Seiberg-Witten transformation and the target space rotation. We have
checked the non-trivial equivalence of the three different
descriptions, with use of a simple example of the non-commutative BPS
Dirac monopole, on the level of not only the solution but also the BPS
equation. We summarize our results in the following table. 

\begin{table}[h]
\begin{center}
\begin{tabular}{c||c|c|c}
description & {\bf (I)} & {\bf (II)} & {\bf (III)} \\
\hline\hline
scalar field  &   $\widehat{\Phi}(x)$ & $\Phi(x)$ &
$\check{\Phi}(\check{x})$ \\ 
\hline
BPS equation  & from linear susy & from linear $+$ non-linear
susy & from linear susy \\ 
\hline
brane & cannot be & tilted D-string & vertical D-string
\\
configuration & understood & \& horizontal D3 & \& tilted D3
\end{tabular}
\begin{minipage}{120mm}
\caption{Three different but equivalent descriptions.}
\end{minipage}
\end{center}
\end{table}

It would be interesting if the calculation performed in this paper can
be extended to the all order in the perturbative expansion in $\theta$
and $\alpha'$. Though we 
have considered the region $r \gg \sqrt{\alpha'} \gg \sqrt{\theta}$ in
this paper, it is expected that the BPS solution considered in this
paper is also a solution of the equations of motion of all order in
$\alpha'$. This expectation follows from the proof in the ordinary
commutative case \cite{Thorlacius:1997bi}. The approach using solitons 
has advantages when one wants to study the equivalence between
non-commutative and commutative descriptions beyond the
perturbation. 

The extension of our analysis to the non-Abelian case is also
important. If the non-linearly realized supertransformation
of the non-Abelian DBI theory is available, then our strategy
can be applied to the non-commutative $U(2)$ monopole whose
commutative description has been briefly considered in sec.\
\ref{sec:u2}. Then it can be shown that the configuration calculated
in sec.\ \ref{sec:u2} subjects to some BPS equations.

The meaning of the target space rotation adopted in sec.\
\ref{sec:rot} is still vague. We have seen 
in sec.\ \ref{sec:repro} that the BPS equations are related with each
other by this rotation. It is to be clarified how this non-trivial
rotation which mixes the fields and the worldvolume coordinates are
consistent with the lagrangian formalism of the DBI action. 

Our final comment is on the `non-commutative eigenvalue
equation'. This shows a correct D-string configuration at least
asymptotically. However, we insist that the eigenvalues to be examined 
are of the commutative description. Since these two apparently
different methods give the same result, some relations must exist
between them. The study of this may provide interesting information of
the non-commutative theory.


\section*{Note added} 
\vs{-5mm}
While this work was in the final stage, we became aware of the paper
\cite{Mateos:2000nc} which shows an overlapping results.
   

\section*{Acknowledgments}
\vs{-5mm}
We would like to thank T.\ Asakawa, H.\ Hata, I.\ Kishimoto and S.\
Moriyama for comments. A part of sec.\ \ref{sec:rot} was
motivated by the discussion with A.\ Hashimoto, and K.\ H.\ would like
to thank him much. This work is supported in part by Grant-in-Aid
for Scientific Research from Ministry of Education, Science, Sports
and Culture of Japan (\#3159, \#3160), and by the Japan Society for
the Promotion of Science under the Predoctoral Research Program. 


\appendix
\section{Non-commutative $1/4$ BPS dyon and Seiberg-Witten
  transformation}  

As studied in sec.\ \ref{sec:u2}, the non-commutative monopoles can be
interpreted in the brane description well by the commutative description
after being performed the Seiberg-Witten transformation
(\ref{eq:phisw}). In this appendix, we investigate the NC $U(3)$
monopoles and the NC $1/4$ dyons in a similar manner.

\subsection{Non-commutative $U(3)$ monopole}

In ref.\ \cite{Hata:2000sj}, a solution for the NC $U(3)$ monopole was
obtained. After performing the Seiberg-Witten transformation to the
solution of ref.\ \cite{Hata:2000sj}, we
obtain the following configuration of the scalar field $Y$:
\begin{eqnarray}
  &&Y=  Y^{(0)}+  Y^{(1)} + {\cal O}(\theta^2), \\
  &&  Y^{(0)}=-\hat{x}_i T_i H(\xi)/r, \\
  &&  Y^{(1)}=\frac{1}{r^3}\biggl(\theta_i \hat{x}_i T_0 U(\xi)
  + \theta_i \hat{x}_j T_{ij} V(\xi)
  +\theta_i \hat{x}_i \hat{x}_j \hat{x}_k T_{jk} W(\xi)
  \biggr),
\end{eqnarray}
where we have defined $\xi \equiv C r$, and all of the conventions
follow from the ones adopted ref.\ \cite{Hata:2000sj}. The functions
$U$, $V$ and $W$ which specify the solution reads (after the
Seiberg-Witten transformation)
\begin{eqnarray}
  &&  U(\xi) = U_{\rm HM}(\xi) +\frac16 H (1-K)(1+K), \nn\\
  &&  V(\xi) = V_{\rm HM}(\xi) +\Half (1-K)(K^2-1), 
  \label{solvcom}\\
  &&  W(\xi) = W_{\rm HM}(\xi) -\frac14 (1-K)(H-2+2K^2+HK). \nn
\end{eqnarray}
In this expression, the functions with ``HM'' mean the ones obtained
in ref.\ \cite{Hata:2000sj}, which give the solution of the NC BPS
equations. The other terms in the right hand sides are produced from
the Seiberg-Witten transformation. It is straightforward to evaluate
the three eigenvalues of the scalar $Y$ as 
\begin{eqnarray}
  \lambda_Y =  \quad
  \frac1{r^3}\theta_i \hat{x}_i
  \left(
    4U-\frac43(V+W)
  \right),\quad
  -(\pm 1)\frac{H}{r} + 
  \frac1{r^3}\theta_i \hat{x}_i
  \left(
    4U+\frac23(V+W)
  \right),
\end{eqnarray}
where $\theta_i\equiv \epsilon_{ijk}\theta_{jk}/2$.
Asymptotically, these become
\begin{eqnarray}
  \lambda_Y =\quad
  \frac{\theta_i\hat{x}_i}{4r^3}
  (1+4z)\xi+{\cal O}(\theta^2)~,\quad
  \mp C \pm \frac1r 
  +\frac{\theta_i\hat{x}_i}{4r^3}
  \left(
    (3-4z)\xi -4
  \right)+{\cal O}(\theta^2).
  \label{eigenc}
\end{eqnarray}
Here the parameter $z$ is included in $V_{\rm HM}$ and $W_{\rm HM}$,
and it indicates a moduli of the relative location of
the two D-strings which suspend between the three D3-branes.
Remarkably, this (\ref{eigenc}) is the same as the result of ref.\
\cite{Hata:2000sj} in which three eigenvalues are obtained by solving
the `non-commutative eigenvalue equation' in the non-commutative
space. Therefore, asymptotically, we obtain the same configuration of
slanted D-string as ref.\ \cite{Hata:2000sj}. However, as in the case
of the NC  
$U(2)$ monopole in sec.\ \ref{sec:u2}, we have the agreement only in
the asymptotic region. Our eigenvalues are regular even at the origin
$r=0$. What is more interesting, the latter two eigenvalues in
(\ref{eigenc}) are arranged to the first order
in $\theta$ in the following way:
\begin{eqnarray}
  \lambda_Y = - (\pm 1)\frac{
    H\Bigl(C(x_i + \theta_i\lambda_Y)\Bigr)}
      {|x_i + \theta_i \lambda_Y|},
  \label{rely}
  \label{eq:lamfin}
\end{eqnarray}
where we have chosen a special value $z = -1/4$ in that case the two
D-strings are aligned\footnote{With this special value of $z$, another
  eigenvalue vanishes exactly.}. This relation (\ref{rely}) indicates
that the eigenvalues (\ref{eigenc}) really exhibits the tilted
D-string configuration with the center on a straight line 
\begin{eqnarray}
  x_i = \theta_i \lambda.
  \label{eq:xtl}
\end{eqnarray}


\subsection{String junction in $B$-field and non-commutative $1/4$ BPS
  dyon}

In the previous subsection, we have obtained a consistent brane
picture of the NC $U(3)$ monopole, 
then let us proceed to the case of the NC $1/4$ BPS dyon studied in
ref.\ \cite{Hata:2000sj}. The authors of ref.\ \cite{Hata:2000sj}
solved the NC Gauss law for another scalar $X$ in the background of
the NC $U(3)$ monopole. We perform the Seiberg-Witten transformation
and obtain the configuration for $X$ in the commutative description as 
\begin{eqnarray}
  &&X=  X^{(0)}+  X^{(1)} + {\cal O}(\theta^2), \\
  &&  X^{(0)}=\frac1r \hat{x}_i \hat{x}_j T_{ij}\frac{Q(\xi)}{\xi}, 
  \\
  &&  X^{(1)}=\frac{1}{\xi r^3}\biggl(\theta_i T_i R(\xi)
  +\theta_i \hat{x}_i \hat{x}_j T_j S(\xi)
  \biggr).
\end{eqnarray}
We choose $(\alpha, \beta)=(0,1)$ and the zero-th order solution is
specified by $Q(\xi) = -2H^2-H+1-K^2$. The functions appearing in the
above are given as  
\begin{eqnarray}
  &&  R(\xi) = R_{\rm HM}(\xi) +\frac13(1-K)(2Q-{\cal D} Q ), 
  \label{solR}\\
  &&  S(\xi) = S_{\rm HM}(\xi) -\frac13(1-K)({\cal D} Q-(5+3K)Q ). 
  \label{solS}
\end{eqnarray}

When $X$ commutes with $Y$, these are simultaneously diagonalizable,
hence the brane interpretation is possible. This requirement provides
a condition 
\begin{eqnarray}
  [X,Y] = \frac{i}{\xi r^3}\theta_k \hat{x}_i \epsilon_{ikm} T_m 
  \Bigl(2 V(\xi)  Q(\xi) + R(\xi) H(\xi)\Bigr) +{\cal O}(\theta^2)=0.
\end{eqnarray}
Interestingly enough, using the solution (\ref{solvcom}), (\ref{solR})
and (\ref{solS}), this condition is satisfied only when $z = -1/4$, in 
whose case the junction interpretation was possible in ref.\
\cite{Hata:2000sj}. 

We calculate three eigenvalues of the scalar field $X$ with $z=-1/4$
as  
\begin{eqnarray}
  \lambda_X =\quad
  \frac83 C - \frac4r+{\cal O}(\theta^2),\quad
  -\frac43 C+ \frac2r \pm\frac{\theta_i\hat{x}_i}{r^3}
  \left(
    2\xi -2
  \right)+{\cal O}(\theta^2),
\end{eqnarray}
where we only write the asymptotic expression.
Again, this shows a perfect agreement with the result from `NC
eigenvalue equation' in ref.\ \cite{Hata:2000sj}. Since the result of
ref.\ \cite{Hata:2000sj} presents a consistent string junction in the
$B$-field whose brane configuration was previously studied in ref.
\cite{Hashimoto:1999}, our commutative picture provides a
consistent configuration of the string junction, too. 

As in the case of $Y$ (\ref{eq:lamfin}), it is possible to arrange the
latter two eigenvalues of $X$ at finite $x$ into the form (to the
first order in $\theta$)
\begin{eqnarray}
  \lambda_X = \frac{2Q\Bigl(C(x_i + \theta_i h(\lambda_X))\Bigr)}
  {3C\Bigl|x_i + \theta_i h(\lambda_X)\Bigl|^2}.
  \label{eq:lamxfin}
\end{eqnarray}
Here the function $h(\lambda_X)$ is defined as
\begin{eqnarray}
  h(\lambda_X)=\mp \frac{H(Cs)}{s}\Biggm|_{s=s(\lambda_X)},
\end{eqnarray}
where the parameter $s$ is a solution of the equation
\begin{eqnarray}
  \lambda_X = \frac{2 Q(Cs)}{3 C s^2}.
\end{eqnarray}
The expression (\ref{eq:lamxfin}) indicates that the $(p,q)$-string
locates on the line 
\begin{eqnarray}
  x_i + \theta_i h(\lambda_X)=0.
  \label{eq:xth}
\end{eqnarray}

\begin{figure}[tdp]
\begin{center}
\leavevmode
\epsfxsize=100mm
\put(225,95){$\theta$}
\epsfbox{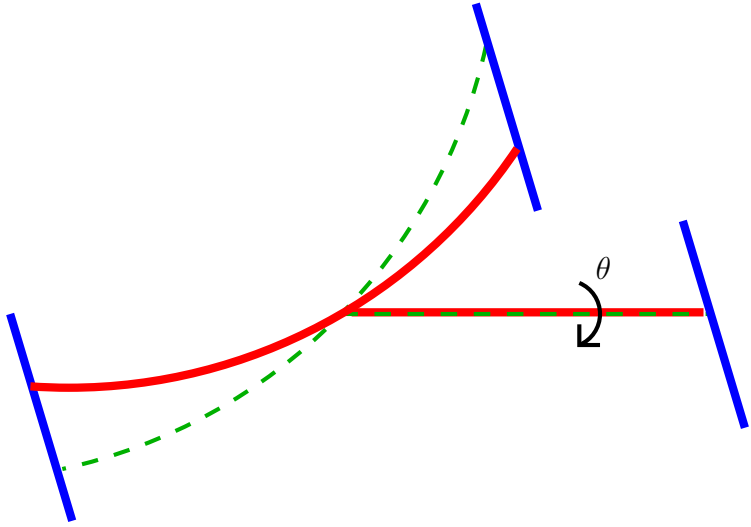}
\caption{The slanted string junction. The dashed line denotes the junction
  of $\theta=0$. Three parallel straight lines are D3-branes. } 
\end{center}
\end{figure}

The whole configuration of the NC string junction is given by the two 
equations (\ref{eq:xtl}) and (\ref{eq:xth}). Eliminating $x_i$ from
these two equations, then we obtain a relation between $\lambda_Y$ and
$\lambda_X$. It is easy to see that this relation is precisely the
same as the one obtained in the usual commutative case
($\theta=0$) of refs.\ \cite{hashimoto:1998sj,hashimoto:1998mp}. 
In ref.\ \cite{hashimoto:1998mp},  the bending of the $(p,q)$-strings
in the network was analyzed, and it was found that the bend of the
strings are consistent with the effective charge defined at a finite
distance $r$. Therefore, we conclude that the bending of the NC string
junction is interpreted in the same way. The difference between the NC
string junction and the previous usual string junction
\cite{hashimoto:1998sj,hashimoto:1998mp} is only that
now the junction is on a plane (\ref{eq:xtl}) tilted by the angle
$\theta$. This fact was predicted in ref.\ \cite{Hashimoto:1999}, and
we have given the proof of the prediction.

\newcommand{\J}[4]{{\sl #1} {\bf #2} (#3) #4}
\newcommand{\andJ}[3]{{\bf #1} (#2) #3}
\newcommand{\AP}{Ann.\ Phys.\ (N.Y.)}
\newcommand{\MPL}{Mod.\ Phys.\ Lett.}
\newcommand{\NP}{Nucl.\ Phys.}
\newcommand{\PL}{Phys.\ Lett.}
\newcommand{\PR}{ Phys.\ Rev.}
\newcommand{\PRL}{Phys.\ Rev.\ Lett.}
\newcommand{\PTP}{Prog.\ Theor.\ Phys.}
\newcommand{\hep}[1]{{\tt hep-th/{#1}}}


\begin{thebibliography}{10}

\bibitem{Hanany:1996ti}
   A.\ Hanany and E.\ Witten,
   {\sl ``Type IIB Superstrings, BPS Monopoles, And Three-Dimensional
   Gauge Dynamics''}, 
   \J{\NP}{B492}{1997}{152}, {\tt hep-th/9611230}.

\bibitem{Witten:1997so}
  E. Witten,
  {\sl ``Solutions Of Four-Dimensional Field Theories Via M Theory''},
  \J{\NP}{B500}{1997}{3}, {\tt hep-th/9703166}.

\bibitem{Maldacena:1997tl}
  J.\ M.\ Maldacena,
  {\sl ``The Large N Limit of Superconformal Field Theories and
  Supergravity''}, 
  \J{Adv.\ Theor.\ Math.\ Phys.}{2}{1998}{231}, 
  \J{Int.\ J.\ Theor.\ Phys.}{38}{1999}{1113}, {\tt hep-th/9711200}.

\bibitem{Leigh:1989}
  R.\ G.\ Leigh,
  {\sl ``Dirac-Born-Infeld action from Dirichlet $\sigma$-model''},
  \J{\MPL}{28}{1989}{2762}.

\bibitem{Hull:1994}
 C.\ M.\ Hull and P.\ K.\ Townsend,
 {\sl ``Unity of Superstring Dualities''},
  \J{\NP}{B438}{1995}{109}, {\tt hep-th/9410167}.

\bibitem{Connes:1998cr}
A.\ Connes, M.\ R. Douglas, and A.\ Schwarz, 
{\sl ``Noncommutative geometry and matrix theory: Compactification on
  tori''},  
  {\em JHEP} {\bf 02} (1998) 003,  
  {\tt hep-th/9711162}.

\bibitem{Douglas:1998fm}
  M.\ R. Douglas and C.\ Hull, 
  {\sl ``D-branes and the noncommutative torus''},
  {\em JHEP} {\bf 02} (1998) 008, 
  {\tt hep-th/9711165}.

\bibitem{Seiberg:1999vs}
  N.\ Seiberg and E.\ Witten, 
  {\sl ``String theory and noncommutative geometry''}, 
  {\em JHEP} {\bf 09} (1999) 032,
  {\tt hep-th/9908142}.

\bibitem{Hashimoto:1999}
  A.\ Hashimoto and K.\ Hashimoto, 
  {\sl ``Monopoles and dyons in non-commutative geometry''},
  {\tt hep-th/9909202}.

\bibitem{Hashimoto:1999bc}
  K.\ Hashimoto, H.\ Hata and S.\ Moriyama,
  {\sl ``Brane Configuration from Monopole Solution in Non-Commutative
    Super Yang-Mills Theory''},
  \J{JHEP}{9912}{1999}{021}, 
  {\tt hep-th/9910196}.

\bibitem{Bak:1999}
  D.\ Bak, 
  {\sl ``Deformed Nahm equation and a noncommutative BPS monopole''},\\
  {\tt hep-th/9910135}.

\bibitem{Hata:2000sj}
   H.\ Hata and S.\ Moriyama,
   {\sl ``String Junction from Non-Commutative Super Yang-Mills
   Theory''}, 
   {\tt hep-th/0001135}.

\bibitem{Callan:1997}
  C.\ G.\ Callan Jr.\ and J.\ M.\ Maldacena, 
  {\sl ``Brane dynamics from the Born-Infeld action''}, 
  {\em Nucl. Phys.} {\bf B513} (1998) 198,
  {\tt hep-th/9708147}.

\bibitem{Hashimoto:1999bi}
  K.\ Hashimoto, 
  {\sl ``Born-Infeld Dynamics in Uniform Electric Field''}, 
  \J{JHEP}{9907}{1999}{016}, {\tt hep-th/9905162}.

\bibitem{Marino:1999ni}
  M.\ Marino, R.\ Minasian, G.\ Moore and A.\ Strominger, 
  {\sl``Nonlinear Instantons from Supersymmetric p-Branes''},
  \J{JHEP}{0001}{2000}{005}, {\tt hep-th/9911206}.

\bibitem{Terashima:1999ui}
  S.\ Terashima, {\sl``U(1) Instanton in Born-Infeld Action and
  Noncommutative Gauge Theory''}, 
  {\tt hep-th/9911245}.

\bibitem{Nekrasov:1998io}
  N.\ Nekrasov and A.\ Schwarz,
  {\sl``Instantons on noncommutative $R^4$, and (2,0) superconformal
  six dimensional theory''}, 
  \J{Commun.\ Math.\ Phys.}{198}{1998}{689}, {\tt hep-th/9802068}.


\bibitem{Braden:1999st}
  H.\ W.\ Braden and N.\ A.\ Nekrasov,
  {\sl``Space-Time Foam From Non-Commutative Instantons''},
  {\tt hep-th/9912019}.

\bibitem{Furuuchi:1999io}
  K.\ Furuuchi,
  {\sl``Instantons on Noncommutative $R^4$ and Projection
    Operators''},\\ 
  {\tt hep-th/9912047}.

\bibitem{dbi} 
  M.\ Born and L.\ Infeld, 
  {\sl ``Foundation of New Field Theory''}, 
  \J{Proc.\ Roy.\ Soc.\ London}{A14}{1934}{425}.

\bibitem{ketov}
  S.\ V.\ Ketov,
  {\sl ``A manifestly $N=2$ supersymmetric Born-Infeld action''},
  \J{Mod.\ Phys.\ Lett.}{A14}{1999}{501}, {\tt hep-th/9809121}.

\bibitem{tsey}
  A.\ A.\ Tseytlin,
  {\sl ``Born-Infeld action, supersymmetry and string theory''},\\
  {\tt hep-th/9908105}.

\bibitem{bagger}
  J.\ Bagger and A.\ Galperin, 
  {\sl ``New Goldstone multiplet for partially broken
  supersymmetry''}, 
  \J{\PR}{D55}{1997}{1091}, {\tt hep-th/9608177}.

\bibitem{Thorlacius:1997bi}
  L.\ Thorlacius,
  {\sl ``Born-Infeld String as a Boundary Conformal Field Theory''},
  \J{\PRL}{80}{1998}{1588}, {\tt hep-th/9710181}.

\bibitem{Mateos:2000nc}
  D.\ Mateos, 
  {\sl``Non-commutative vs. Commutative Descriptions of D-brane BIons''},
  {\tt hep-th/0002020}.

\bibitem{hashimoto:1998sj}
  K.\ Hashimoto, H.\ Hata and N.\ Sasakura,
  {\sl ``3-String Junction and BPS Saturated Solutions in SU(3)
    Supersymmetric Yang-Mills Theory''},  
  \J{\PL}{B431}{1998}{303}, {\tt hep-th/9803127}.

\bibitem{hashimoto:1998mp}
  K.\ Hashimoto, H.\ Hata and N.\ Sasakura,
  {\sl ``Multi-Pronged Strings and BPS Saturated Solutions in SU(N)
    Supersymmetric Yang-Mills Theory''},
  \J{\NP}{B535}{1998}{83}, {\tt hep-th/9804164}.

\end{thebibliography}
\end{document}